\documentclass[prd,showpacs,nofootinbib,preprintnumbers,preprint,floatfix,superscriptaddress]{revtex4}

\usepackage{amsmath}
\usepackage{amssymb}
\usepackage{graphicx}
\usepackage{dsfont}

\usepackage[dvipsnames]{xcolor}

\makeatother

\begin{document}

\title{Topological defects in superconductor open nanotubes under gradual and abrupt switch-on of the transport current and magnetic field}

\author{I. \surname{Bogush}}
\email{igbogush@gmail.com}
\affiliation{Institute for Integrative Nanosciences, Leibniz IFW Dresden, Helmholtzstraße 20, D-01069 Dresden, Germany}
\affiliation{Faculty of Physics, Lomonosov Moscow State University, 119899, Moscow, Russia}
\affiliation{Moldova State University, strada A. Mateevici 60, MD-2009 Chisinau, Republic of Moldova}
\author{V. M. \surname{Fomin}}
\email{v.fomin@ifw-dresden.de}
\affiliation{Institute for Integrative Nanosciences, Leibniz IFW Dresden, Helmholtzstraße 20, D-01069 Dresden, Germany}
\affiliation{Moldova State University, strada A. Mateevici 60, MD-2009 Chisinau, Republic of Moldova}
\affiliation{Institute of Engineering Physics for Biomedicine, National Research Nuclear University “MEPhI”, Kashirskoe shosse 31, 115409 Moscow, Russia}

\begin{abstract}
We analyze dynamics of the order parameter in superconductor open nanotubes under a strong transport current in an external homogeneous magnetic field using the time-dependent Ginzburg-Landau equation. Near the critical transport current, the dissipation processes are driven by vortex and phase-slip dynamics. A transition between the vortex and phase-slip regimes is found to depend on the external magnetic field only weakly if the magnetic field and/or transport current are switched on gradually. In the case of an abrupt switch-on of the magnetic field or transport current, the system can be triggered to the stable phase-slip regime within a certain window of parameters. Finally, a hysteresis effect in the current-voltage characteristics is predicted in superconductor open nanotubes. 
\end{abstract}


\maketitle

\section{Introduction}
Topological objects are indestructible by smooth transformations, and thus their stability is protected by topological considerations \cite{Eschrig:2011}. Topological defects in a superconductor turn the order parameter to zero and may lead to the emergence of a finite resistance. The well-known topological defects are {\it vortices} in a bulk superconductor. The centerline of a vortex is called a vortex core, where the order parameter amplitude is zero and its phase is not defined. Encircling the vortex core over an arbitrary path gives a phase change $2\pi n$ quantized by an integer winding number $n\in\mathbb{Z}$.

The concept of the {\it phase slippage} was introduced for the resistive state of the narrow quasi-1D superconductor filaments in Ref. \cite{Langer:1967}. At the phase-slip event in a 1D nanowire, the order parameter vanishes at some point, and the phase suffers a jump equal to $2\pi$. There are various mechanisms of the phase-slip appearance, e.g. Langer-Ambegaokar-McCumber-Halperin (LAMCH) mechanism of thermal-dominated fluctuation-driven regime near the critical temperature \cite{Langer:1967, Tidecks:1990}, inhomogeneity-driven phase slip \cite{Skocpol:1974}, quantum phase slip \cite{Giordano:1988}.

Quantum phase slip in a 2D superconductor has been recently found through magnetotransport measurements \cite{Saito:2018}. The 2D mechanism of phase-slip emergence related to unbinding of vortex-antivortex pairs below the Berezinski-Kosterlitz-Thouless (BKT) transition was developed in Refs. \cite{Berezinskii:1971-1,Berezinskii:1971-2,Kosterlitz:1973}. For nanowires of 100 nm, it was shown \cite{Bell:2007-1, Bell:2007-2} that LAMCH-mechanism \cite{Langer:1967,McCumber:1970} dominates over BKT at relatively low current. In the interval of temperatures between the BKT critical temperature and the critical temperature, thermal fluctuations are sufficient to unbind vortex-antivortex pairs \cite{Halperin:1979}. The BKT scenario was induced by a magnetic field in a 2D spin-dimer system with a multilayer magnet \cite{Tutsch:2014}. The BKT effect was found in a trapped quantum degenerate gas of rubidium atoms \cite{Hadzibabic:2006}. Another type of phase transition found experimentally in a 2D superconducting vortex system is a liquid-solid transition \cite{Chen:2007}. Bose-Einstein condensate in a 3D optical lattice provides an experimental evidence for the temperature-independent dissipation \cite{McKay:2008} as well as an ultracold quantum gas in 1D optical lattice does for the velocity-dependent dissipation \cite{Tanzi:2006}. The crossover between thermal and quantum phase slips controlled by velocity was detected in 1D superfluid tubes \cite{Scaffidi:2017}.

If the transport current is strong enough, the energy barrier, which binds the vortex-antivortex pair, can be overcome, and thermally induced resistivity should be dominant. In an external magnetic field, vortices move due to the Magnus force induced by transport current \cite{Thinkham:1996} and contribute to resistivity. Even in a weak magnetic field, dissipation due to vortex movement dominates over the thermally activated phase-slip events in superconductor submicron wires \cite{Bell:2007-1}.

Novel superconducting nanostructured microarchitectures, e.g., open nanotubes \cite{Fomin:2012,Rezaev:2019,Dobrovolskiy:2021} and nanocoils \cite{Fomin:2017,Losch:2019,Cordoba:2019}, provide new opportunities due to their nontrivial geometry. If a nanotube is thin enough, only the component of the magnetic field normal to its surface plays a role in the order-parameter dynamics. This suggests making use of geometry to manipulate the effective magnetic field profile. As a result, the distribution of the order parameter is highly inhomogeneous, which opens the way to an interplay and transitions between vortex and phase-slip regimes. Fingerprints of the vortex and phase-slip patterns were found in nanohelices 100 nm in diameter experimentally, supported by numerical simulations based on the time-dependent Ginzburg-Landau (TDGL) equation \cite{Cordoba:2019}. In Ref. \cite{Fomin:2020}, a transition between different vortex and phase-slip patterns as a function of the transport current density and the external magnetic field was revealed for niobium and tin open nanotubes with 400 nm radius. The nontrivial profile of the magnetic field in open nanotubes introduces new features in the superconducting dynamics. For example, one can distinguish the superconducting regime with two channels of the suppressed superconductivity along the nanotube with moving vortices inside, which cannot be found in a planar nanomembrane. Vortices in both semi-cylinders are moving in opposite directions (being unidirectional in the unfolded planar case) providing a new feature for studies. A planar nanostructure with a region of suppressed superconductivity contains one or a few chains of vortices moving in the same direction, while in a nanotube with phase-slip, there are nucleating and annihilating vortex-antivortex pairs moving towards each other. The transition between Abrikosov-Josephson vortex and phase-slip regimes was reported in Josephson junction depending on the junction length and transport current \cite{Sheikhzada:2017}. However, the mechanism of the phase-slip appearance is claimed to be attributed to the effective non-locality in the order parameter dynamics, whilst the phase-slip in nanotubes is obtained without non-locality effects \cite{Fomin:2020}.

The hysteresis effects in superconductor magnetization resulting from the surface currents, flux pinning and existence of different phases are well known \cite{Schweitzer:1967-1, Schweitzer:1967-2, Schweitzer:1968-3}. The hysteresis effects in current-voltage characteristics are found theoretically and experimentally in superconductor nanowires and microbridges \cite{Tidecks:1990, Vodolazov:2004, Vodolazov:2011, Segev:2011, Likharev:1979}. The explanation of the hysteresis effect is attributed to the change of the effective temperature of quasiparticles due to the Joule heating \cite{Gubankov:1972,Skocpol:1974-2,Tinkham:2003,Hazra:2010} or the finite relaxation time of the order parameter magnitude \cite{Likharev:1975,Song:1976,Hojgaard:1976,Dover:1981,Baratoff:1977,Lozanne:1986,Michotte:2004}. The hysteresis effect attributed to the phase-slip regime was obtained for a mesoscopic superconductor square with attached contacts \cite{Vodolazov:2005} and NbN superconductor nanowires \cite{Elmurodov:2008}. However, the hysteresis effect obtained in these two systems appears due to the coupling of the order parameter dynamics with the heat equation. When the system triggers the phase-slip regime, the normal current becomes larger, leading to the higher dissipation power and higher temperature of the sample. In its turn, the higher temperature of the sample favors the phase-slip regime, making the hysteresis effect appreciable. The hysteresis effect found in this paper appears purely dynamically (due to the energy barrier between different superconducting regimes) at the constant temperature (with no coupling to the heat equation).

In the present paper, we push forward the idea of Ref. \cite{Fomin:2020}, studying superconductivity in open nanotubes and consider transitions between superconducting regimes. While in Ref. \cite{Fomin:2020}, the superconductor nanotubes were shown to demonstrate non-trivial behavior in dependence of the external parameters (transport current density $j_{tr}$ and magnetic field $B$), we pay attention to the dependence of the regime on how the superconducting state was prepared, i.e. the way of switch-on of those parameters. The dependency of the superconducting regime on the state preparation can be understood as a memory effect, and it points out the importance of the appropriate system preparation in numerical calculations.

Sec. II represents the model based on the TDGL equations. In Sec. III, the induced voltage is shown to increase mainly monotonically with $j_{tr}$ and $B$ if they are switched on gradually. In Sec. IV, it is demonstrated that the abrupt transport current switch-on can cause a transition to the stable phase-slip regime, which is characterized by a higher induced voltage, than the vortex regime. In Sec. V, a similar effect for the abrupt magnetic field switch-on is found. In Sec. VI, we investigate the hysteresis effect in current-voltage characteristics due to the stability of different regimes (particularly, phase-slip and vortex regimes). Sec. VII contains discussions of the obtained results and outlines their potential application.

\section{Model}
\begin{figure}
    \centering
    \includegraphics[width=0.5\textwidth]{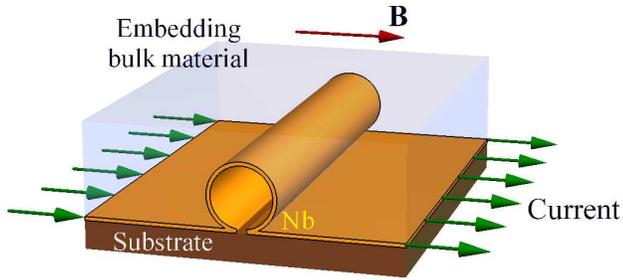}
    \caption{ Scheme of the open nanotube embedded into a heat sink. }
    \label{fig:cylinder}
\end{figure}

A physical model consists of a cylinder of a small finite thickness $d$ with a small paraxial slit (Fig. \ref{fig:cylinder}). The cylinder is embedded in a heat sink, and the slit banks are connected to the contacts carrying a transport current. The cylinder is assumed to be thin enough in order to neglect the finite-thickness effects as discussed below. One of the conditions for applicability of the 2D approximation is that the induced magnetization has a negligible influence on the order parameter in other points (approximately, it can be expressed as $d < \lambda$, where $\lambda$ is the penetration depth). Another applicability condition arises from the fact that the tangent magnetic field can lead to nucleation of vortices with cores across the cylinder thickness. Thus, the cylinder should be thinner than the vortex diameter $\sim4\xi$, where $\xi$ is the coherence length. Also, we will neglect the impact of the induced magnetization of the cylinder. The condition of the applicability of this approximation is the GL parameter $\kappa$ larger than some characteristic value as a function of $d$. Otherwise, the coupling between the electromagnetic field and the order parameter is strong, which modifies the order parameter behavior both quantitatively and qualitatively \cite{Brandt:2005,Brandt:2009}. The smallness of the effects of the induced magnetization and the tangent component of the magnetic field for Nb ($\kappa=4.7$) C-shaped microdevice was demonstrated in Ref. \cite{Smirnova:2020}. When all these conditions are met, the 3D equations of motion for the superconductor sample can be reduced to purely 2D ones.

We consider further an infinitely thin cylinder $D$ of radius $R$, length $L$ with a slit of an arc-length $\delta$. The cylinder allows for a parametrization in two Cartesian coordinates: the arc-length $x\in [\delta/2, 2\pi R - \delta / 2]$ and the length along the cylinder $y\in[0, L]$. Such a cylinder can be bent from a flat membrane with a width $W=2\pi R - \delta$. The cylinder is placed in a homogeneous magnetic field $\mathbf{B} = B \mathbf{e}_z$. The magnetic field component tangent to the cylinder surface does not influence superconducting dynamics, and only the normal component $B_n = (\mathbf{B} \cdot \mathbf{n})$ (with a unit vector $\mathbf{n}$ normal to the cylinder surface) should be taken into consideration.

Following Ref. \cite{Fomin:2020}, we use 2D TDGL dimensionless equation in the external magnetic field (see Tables \ref{tab:params_1}, \ref{tab:scale_1})
\begin{equation} \label{eq:gl}
    (\partial_t + i\kappa \varphi) \psi = \frac{1}{\kappa^2}\left(\boldsymbol{\nabla} - i\kappa \mathbf{A}\right)^2 \psi + (1 - |\psi|^2)\psi,
\end{equation}
where $t$ is time, the vector potential $\mathbf{A}$ describes the magnetic field normal to the cylinder surface ${B_n \mathbf{n} = [\boldsymbol{\nabla} \times \mathbf{A}]}$, the scalar potential $\varphi$ describes the electric field $\mathbf{E}=-\boldsymbol{\nabla}\varphi$, and the complex scalar field $\psi$ is the order parameter. From the continuity equation of the total current density (superconducting + normal) $\boldsymbol{\nabla} \cdot (\mathbf{j}_{sc} + \mathbf{j}_{n}) = 0$, the Poisson equation for the scalar potential $\varphi$ follows
\begin{equation}
    \Delta \varphi = \frac{1}{\sigma} \boldsymbol{\nabla} \cdot \mathbf{j}_{sc},\qquad
    \mathbf{j}_{sc} = \frac{1}{2i\kappa}(\psi^* (\boldsymbol{\nabla} - i\kappa \mathbf{A}) \psi - \text{c.c.}),\qquad
    \mathbf{j}_{n} = \sigma \mathbf{E},
\end{equation}
where $\sigma$ is a normal conductivity. Absence of the induced magnetization in our model allows us to exclude $\kappa$ from the GL and Poisson equations (and the corresponding boundary conditions) through the following transformations: $x \to x / \kappa$, $\varphi \to \varphi / \kappa$, $B \to \kappa B$.  However, we will keep $\kappa$ in our equations for the sake of convenience, but one should remember that all results are fair for other systems meeting the aforementioned conditions of the model applicability, which can be obtained by the above transformations.

The boundary conditions for GL equation have the form \cite{deGennes:1999,Thinkham:1996}
\begin{equation}
    \mathbf{n} \cdot \left( \boldsymbol{\nabla} - i\kappa \mathbf{A} \right)\psi = -\frac{1}{b}\psi,
\end{equation}
where $b\to \infty$ corresponds to an insulator, $b$ approaches $0$ for a magnetic material, and normal metals have a finite $b$. For our purposes, we assume that $b\to\infty$ at free boundaries and $b=0$ at the contacts
\begin{equation}
    \left. \left( \partial_y - i\kappa A_y \right)\psi \right|_{\partial D_y}= 0,\qquad
    \left.\psi\right|_{\partial D_x} = 0. 
\end{equation}
The transport current density $\mathbf{j}_{tr}$ is introduced through the boundary conditions on the scalar potential
\begin{equation}\label{eq:phi_bc}
    \left. \partial_y \varphi \right|_{\partial D_y} = 0,\qquad
    \left. \partial_x \varphi \right|_{\partial D_x}  = - j_{tr}/\sigma.
\end{equation}

\begin{table}[ht] \caption[]{Materials and geometric parameters used for the simulations\footnote{See Ref. \cite{Fomin:2020} for details; the dirty limit is used}.}\label{tab:params_1}
\begin{center}\begin{tabular}{ |c|c|c| }
 \hline
 & Denotation & Value for Nb\\
 \hline\hline
 Relative temperature & $T/T_c$ & 0.95/0.952/0.955
 \\
 Penetration depth & $\lambda=\lambda_0\sqrt{\xi_0 / (2.66(1-T/T_c))}$ & 273/278/287 nm
 \\
 Coherence length & $\xi=0.855\sqrt{\xi_0 l / (1-T/T_c)}$ & 58/60/62 nm
 \\
 GL parameter & $\kappa=\lambda/\xi$ & 4.7
 \\
 Fermi velocity & $v_F=\sqrt{2E_F/m_e}$ & $6\times 10^{-5}$ m/s
 \\
 Thickness of the film & $d$ & 50 nm
 \\
 Mean free electron path & $l$ & 6.0 nm
 \\
 Diffusion coefficient & $D=lv_F/3$ & $1.2 \times 10^{-3} \text{ m}^2/\text{s}$
 \\
 Normal conductivity for a thin membrane & $\sigma=l/[3.72\times 10^{-16}\;\Omega\,\text{m}^{2}]$ & 16 $(\mu\Omega \text{ m})^{-1}$
 \\
 Cylinder radius & $R$ & 390 nm
 \\
 Length & $L$ & 5 $\mu$m
 \\
 \hline
\end{tabular}\end{center}
\end{table}

\begin{table}[ht]  \caption{Units for the dimensionless quantities. \label{tab:scale_1}}
\begin{center}\begin{tabular}{ |c|c|c| }
 \hline
 & Unit & Value for Nb at $T/T_c = 0.95$ \cite{Fomin:2020}\\
 \hline\hline
 Time & $\xi^2/D$ & 2.8 ps
 \\
 Length & $\lambda$ & 273 nm
 \\
 Magnetic field & $\Phi_0 / 2\pi\lambda\xi$ & 20.6 mT
 \\
 Current density & $\hbar c^2 / 8\pi\lambda^2 \xi e$ & 60 GA $\text{m}^{-2}$
 \\
 Electric potential & $\sqrt{2}H_c \lambda^2 / c \tau $ & 540 $\mu$V
 \\
 Conductivity & $c^2 / 4\pi\kappa^2D$ & 31 $(\mu\Omega \text{ m})^{-1}$
 \\
 \hline
\end{tabular}\end{center}
\end{table}

Integration schemes that are not gauge invariant may introduce large errors in numerical simulations. To avoid this issue, we use the link variables \cite{Kato:1993}
\begin{align}
    &
    U_x^{ba} = \exp\left(
        -i\kappa \int_{x_a}^{x_b} A_x(x', y) dx'
    \right),\qquad
    U_y^{ba} = \exp\left(
        -i\kappa \int_{y_a}^{y_b} A_y(x, y') dy'
    \right),\qquad
    \\\nonumber &
    U_t^{ba} = \exp\left(
        i\kappa \int_{t_a}^{t_b} \varphi(t', x, y) dt'
    \right).
\end{align}
For the purposes of the numerical methods, the integral for the link between two points of the grid can be approximated via a middle-point value
\begin{equation}
    U_x^{ba} \approx \exp\left(
        -i\kappa A_x\left(\frac{x_a+x_b}{2}, y\right) \Delta x
    \right),\qquad
    U_y^{ba} \approx \exp\left(
        -i\kappa A_y\left(x, \frac{y_a+y_b}{2}\right) \Delta y
    \right).
\end{equation}
When solving the Poisson equation, we consider that the speed of light is much higher than the speed of any other processes related to the order parameter dynamics, such as the speed of vortices (light travels along the cylinder length during $\sim0.02\text{ps}$). This allows us to solve the elliptic equation (with Laplace operator) instead of the hyperbolic one (with d'Alembertian), what is numerically simpler and faster. The solution of the Poisson equation provides the value of the scalar potential $\varphi$ exactly at the same instant, when we know the value of the order parameter $\psi$ and the corresponding superconducting current density. So, we have to use this instantaneous value for the numerical link variable
\begin{align}
    U_t^{ab} = \exp\left(
        i\kappa \varphi(t, x, y) \Delta t
    \right)
\end{align}
instead of the middle-point value. Finally, to get the finite-difference scheme, the following rules
\begin{align} \label{eq:der_approximation}
    \left(\partial_x - i \kappa A_x\right) \psi \approx 
    \frac{U^{ab}_x \psi_a - \psi_b}{\Delta x},\qquad
    \left(\partial_x - i \kappa A_x\right)^2 \psi \approx 
    \frac{U^{ab}_x \psi_a - 2\psi_b + U^{cb}_x\psi_c}{(\Delta x)^2}
\end{align}
and similar rules for $y$, $t$ are used. In order to construct an explicit numerical scheme, we use approximations (\ref{eq:der_approximation}) and solve GL equation (\ref{eq:gl}) with respect to the next time step
\begin{equation}
    \psi(t + \Delta t) = (U_t^{t+\Delta t, t})^{-1} \cdot \left(\psi(t) + F[\psi, U_x^{ab}, U_y^{ab}] \cdot \Delta t\right),
\end{equation}
where $F$ is the right-hand side of Eq. (\ref{eq:gl}) approximated through the link variables.

The voltage between the contacts is estimated as an averaged difference of the scalar potentials
\begin{equation}
    U = \frac{1}{L\Delta T} \int_{T_0}^{T_0 + \Delta T} dt \int_0^L dy \left( \varphi(t, W, y) - \varphi(t, 0, y) \right)
\end{equation}
starting from some time $T_0$ during an interval $\Delta T$, which is much larger than any characteristic time of the quasi-periodic stationary processes in the system, such as vortex nucleation period and its time of flight through the cylinder.

For the numerical calculations, we exploit the finite-difference method with a grid containing $192\times384$ points and a step $\Delta t = \text{0.03 ps}$. A set of calculations performed with a finer grid and a shorter time step guarantees that the result is stable with using the link variables. The process is considered stationary and stable, if the system demonstrates a quasi-periodic behavior (in terms of free energy, voltage and visually for patterns of $|\psi|$, $\text{arg}\psi$ and potential) for the time much longer than the characteristic time of the processes mentioned in the previous paragraph.

We solve the Poisson equation using the iterative method after each step of the numerical iteration for the order parameter. Stopping criterion for the iterative solver is the smallness of the residual $\varepsilon = |\Delta \varphi - \frac{1}{\sigma} \boldsymbol{\nabla} \cdot \mathbf{j}_{sc}|$. In addition, the solver performs at least 10 iterations even if the stopping criterion has already been satisfied. In order to optimize the calculations, we choose $\varepsilon < 0.02$ since a further increase of the precision does not change the result quantitatively. We use the previous solution for $\varphi$ as a seed solution for the next time step to accelerate the calculations.

The iterative method for the Poisson equation works similarly to the iterative method for the heat equation, so the residual blurs out in the numerical solution with each iteration. As a result, if the boundary conditions are changed, one should perform as many iterations as is required to spread numerical corrections from one side of the grid to the opposite one. In this case, the number of the required iterations should be of the order of the number of the grid points ($\sim 200$). Since this number is very big, we make use of the standard methods of the partial differential equations to simplify the problem analytically (similarly to Ref. \cite{Sadovskyy:2015}). We split the scalar potential into two parts $\varphi_{dl}$, $\varphi_{ind}$, satisfying the following equations
\begin{equation}
    \Delta\varphi_{dl} = 0,\qquad
    \left. \partial_y \varphi_{dl} \right|_{\partial D_y} = 0,\qquad
    \left. \partial_x \varphi_{dl} \right|_{\partial D_x}  = - j_{tr}/\sigma,
\end{equation}
\begin{equation}
    \Delta\varphi_{ind} = \frac{1}{\sigma}\boldsymbol{\nabla}\cdot\mathbf{j}_{sc},\qquad
    \left. \mathbf{n}\cdot \boldsymbol{\nabla} \varphi_{ind}\right|_{\partial D} = 0.
\end{equation}
The first part $\varphi_{dl}$ represents the normal non-divergent current density, and the second part $\varphi_{ind}$ can be associated with a potential induced by the superconducting current density. The sum of these terms ${\varphi = \varphi_{dl} + \varphi_{ind}}$ gives the solution for the original equation with its boundary conditions. The analytical expression for the first part is
\begin{equation}
    \varphi_{dl} = -j_{tr}(x-\pi R)/\sigma,
\end{equation}
and the only remaining part to be calculated numerically is $\varphi_{ind}$.

The numerical algorithm is implemented in CUDA C++ \cite{cuda_docs} (with a wrapper written in Rust language \cite{rust_docs}) in order to exploit the advantages of the high parallelism of graphical processing units.

\section{Gradual current switch-on}
\label{sec:gradual}

One can na{\"\i}vely expect that the voltage between contacts of the cylinder grows monotonically as a function of the magnetic field and transport current density. In Ref. \cite{Fomin:2020} it was shown that there is a peak of higher voltage for a narrow interval of the external parameters, adding peaks to the function $U(B, j_{tr})$ due to the transitions between the vortex and phase-slip regimes. Those numerical experiments have been performed as follows: prepare an initial random state of the order parameter, switch on the magnetic field and after a relaxation (typically, $\sim 80 \text{ ps}$) switch on the transport current instantly. When the transport current is switched on, the vortex pattern that occurred in the external magnetic field undergoes changes and adapts for the new conditions. If the transport current is switched on too fast, the order parameter may suffer the impact of a high normal current density, and come up with a metastable state containing a phase slip. A metastable state means that the state satisfies equations of motion in a stationary or quasi-stationary regime, but there is another (quasi-) stationary regime with a lower energy. As we will see further, the high-voltage peaks may appear due to the relaxation of the system to a metastable state in the regime of an abrupt switch-on of the transport current. An energy barrier between the phase-slip and the vortex regimes can prevent the system to return to the vortex regime. A similar effect can be achieved when the magnetic field is switched on abruptly after the transport current has been switched on. The time of the (linear) switch-on of the transport current or magnetic field will be called the \textit{ramp time} of the corresponding external parameter, and will be denoted $\Delta t_{c}$, $\Delta t_{f}$ respectively. The switch-on process of $B$ or $j_{tr}$ will be called \textit{ramping}.

The numerical experiment with a gradual switch-on of the current consists of the following steps.
\begin{itemize}
    \item S1. Initialize the system with $\psi = 1 + 5\times 10^{-3}(a + i b)$ and $\varphi = 0$, where $a, b$ are random real numbers distributed normally as $\mathcal{N}(0, 1)$.
    \item S2. Switch on the magnetic field linearly from 0 to $B$ in 10 ps.
    \item S3. Let the system relax in 100 ps.
    \item S4. Switch on the transport current density linearly from $0$ to $j_{tr}$ during $\Delta t_{c} = 500 \;\text{ps}$ (for some calculations we took even longer transport current ramp time $\Delta t_{c}$ up to 10 ns in order to achieve a better convergence to the final stationary state or to provide necessary gradualness of the ramping). Parameter values achieved at this step will be called ``final''.
    \item S5. Let the system come to the final (quasi-)stationary regime during 6 to 9 ns (for some calculations up to 60 ns if the state does not converge to stationary one for a long time).
\end{itemize}

\begin{figure}
    \centering
        \includegraphics[width=0.96\textwidth]{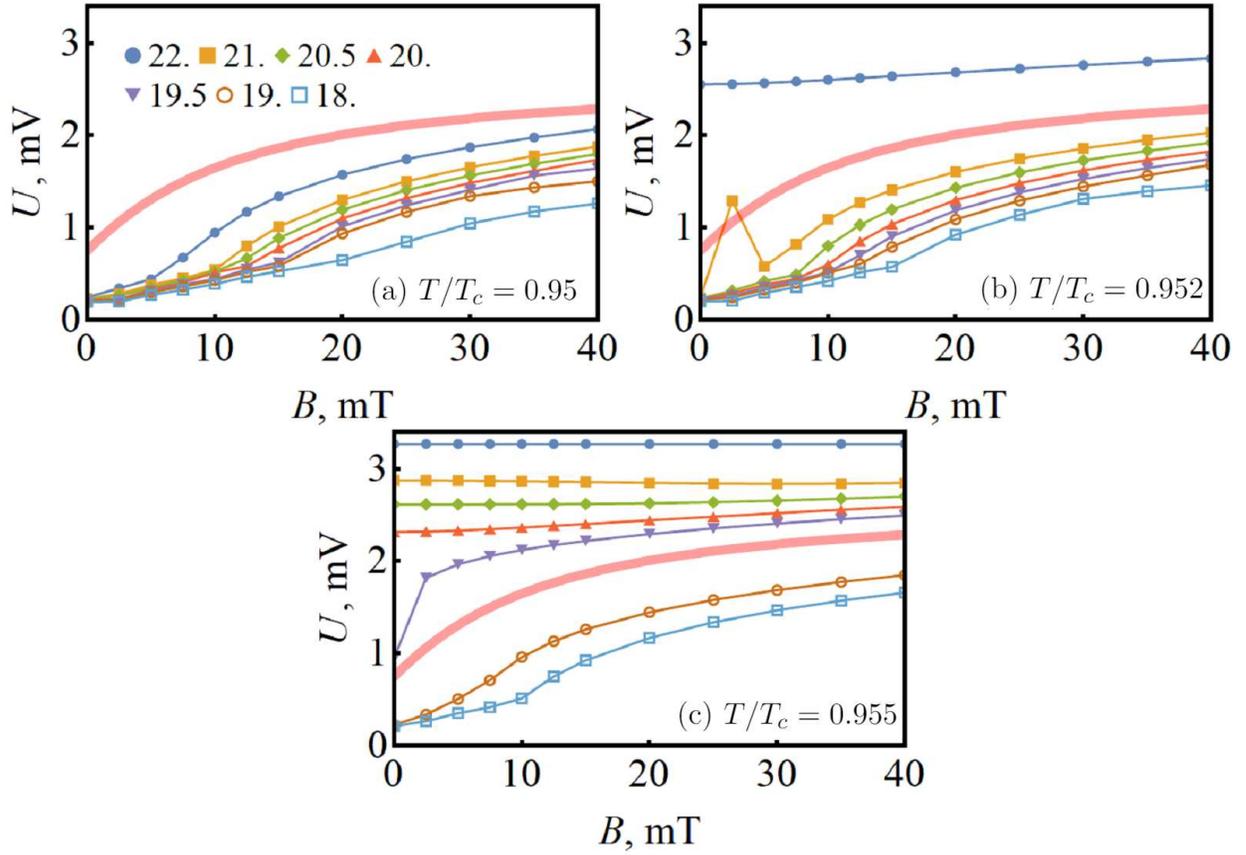}
    \caption{Average voltage induced in Nb open nanotubes with $R=390\;\text{nm}$ and slit $\delta=60 \text{ nm}$ in the external magnetic field $B$ at a gradually switched on transport current density $j_{tr}$. Different curves correspond to different values of $j_{tr}$ (in $\text{GA}/\text{m}^2$). The transport current ramp time is $\Delta t_c = 500 \text{ ps}$ (and longer, for some points up to 10 ns). The pink bold line that separates vortex and phase-slip regimes is drawn manually.}
    \label{fig:adiabatic_U}
\end{figure}

The results for different relative temperatures $T/T_c$ are depicted in Fig. \ref{fig:adiabatic_U}. The transition between the vortex regime with a lower voltage and a phase slip regime with a higher voltage occurs at some value of $j_{tr}$, which decreases with rising temperature. The approximate boundary between the vortex and phase-slip regimes is estimated to be at around 2 mV, except for the case of the weak magnetic fields, when the phase-slip regime is characterized by a lower voltage (e.g. 1 mV for $T/T_c = 0.955,\, j_{tr}= 19.5\text{ GA}/\text{m}^2, \, B=0 \text{ mT}$). The transition to the phase-slip regime does not depend on the magnetic field $B$, except one peak at $T/T_c = 0.952,\, j_{tr}= 21\text{ GA}/\text{m}^2, \, B=2.5 \text{ mT}$, which will be considered at the end of the present section.

\begin{figure}
    \centering
    \includegraphics[width=0.95\textwidth]{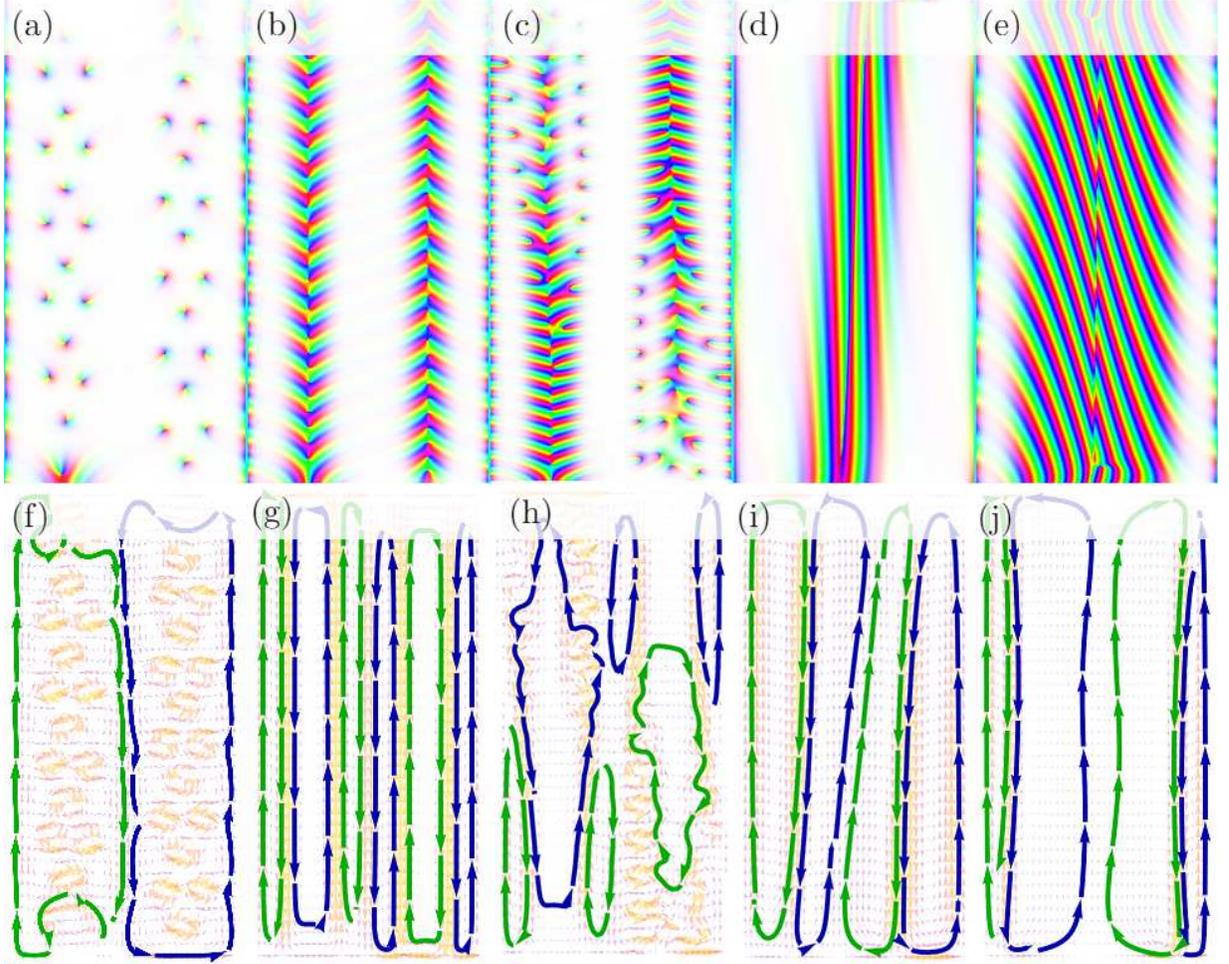}
    \caption{ Examples of the regimes depicted for the order parameter (upper panels) and total screening current density (lower panels) at $T/T_c=0.952$. Each panel represents the unfolded open nanotube with height $L=5\,\mu\text{m}$ and the membrane width $W = 2.39 \,\mu\text{m}$. From left to right: vortex lattice (${j_{tr}=18 \text{ GA/m}^2}, {B=12.5 \text{ mT}}$), dense vortex chain (${j_{tr}=20 \text{ GA/m}^2}, {B=20 \text{ mT}}$), dense vortex chain decaying into separate vortices (${j_{tr}=18 \text{ GA/m}^2}, {B=35 \text{ mT}}$), a weak phase slip (${j_{tr}=21 \text{ GA/m}^2}, {B=2.5 \text{ mT}}$), a strong phase slip (${j_{tr}=22 \text{ GA/m}^2}, {B=10 \text{ mT}}$). Current flows are obtained by numerical integration of the total screening current density. Green (blue) flows are clockwise (counter-clockwise). Orange vectors stand for the vector field of the total screening current density. Color map of complex values is defined in Fig. \ref{fig:complex_map}.}
    \label{fig:adiabatic_regimes}
\end{figure}

Under a low magnetic field and a weak transport current density, vortices and phase slips are not present in the system. However, the current injected from the contacts is not superconducting, which generates the contact voltage (the contact resistance can be estimated from the results at ${B=0 \text{ mT}}$ with purely superconducting state without vortices and phase slips $U/j \approx 11\,\text{k}\Omega\,\text{m}^2$).
Approaching the critical transport current density, and if the magnetic field is high enough to lead to nucleation of vortices, dynamics of the topological defects become nontrivial and richer. There are three main regimes in addition to their intermediate states. The first regime occurs for weak current density and/or low magnetic fields, representing a fragment of {\it vortex lattice} with clearly visible separate vortex cores in each half-cylinder (Fig. \ref{fig:adiabatic_regimes}a, \ref{fig:adiabatic_regimes}f).
The second regime occurs for stronger current density and higher magnetic fields, representing two channels of suppressed superconductivity in each half-cylinder with vortices traveling in each channel forming {\it dense vortex chains} (Fig. \ref{fig:adiabatic_regimes}b, \ref{fig:adiabatic_regimes}g). We will call the vortex pattern sparse if vortices are clearly separated and the amplitude $|\psi|$ forms a core with a radius of the order $\xi$. On the contrary, we will call vortex pattern dense if vortices move inside a region of the suppressed superconductivity. If the magnetic field is increased, such channels with vortex chains can decay into separate vortices travelling in the half-cylinders (Fig. \ref{fig:adiabatic_regimes}c, \ref{fig:adiabatic_regimes}h). 
The third regime consists of a {\it phase slip} in the region of the cylinder opposite to the slit with quickly nucleating, moving and annihilating vortex-antivortex pairs (Fig. \ref{fig:adiabatic_regimes}d, \ref{fig:adiabatic_regimes}i, see \cite{Fomin:2020} for details). For the higher values of the transport current density, the region of the suppressed superconductivity expands practically over the entire cylinder with a phase slip (Fig. \ref{fig:adiabatic_regimes}e, \ref{fig:adiabatic_regimes}j). According to the results of the numerical calculations, a transition from the vortex to the phase-slip regime is accompanied by a voltage jump, therefore the phase-slip regime is characterized by a higher voltage than the vortex regime.

\begin{figure}
    \centering
    \includegraphics[width=0.35\textwidth]{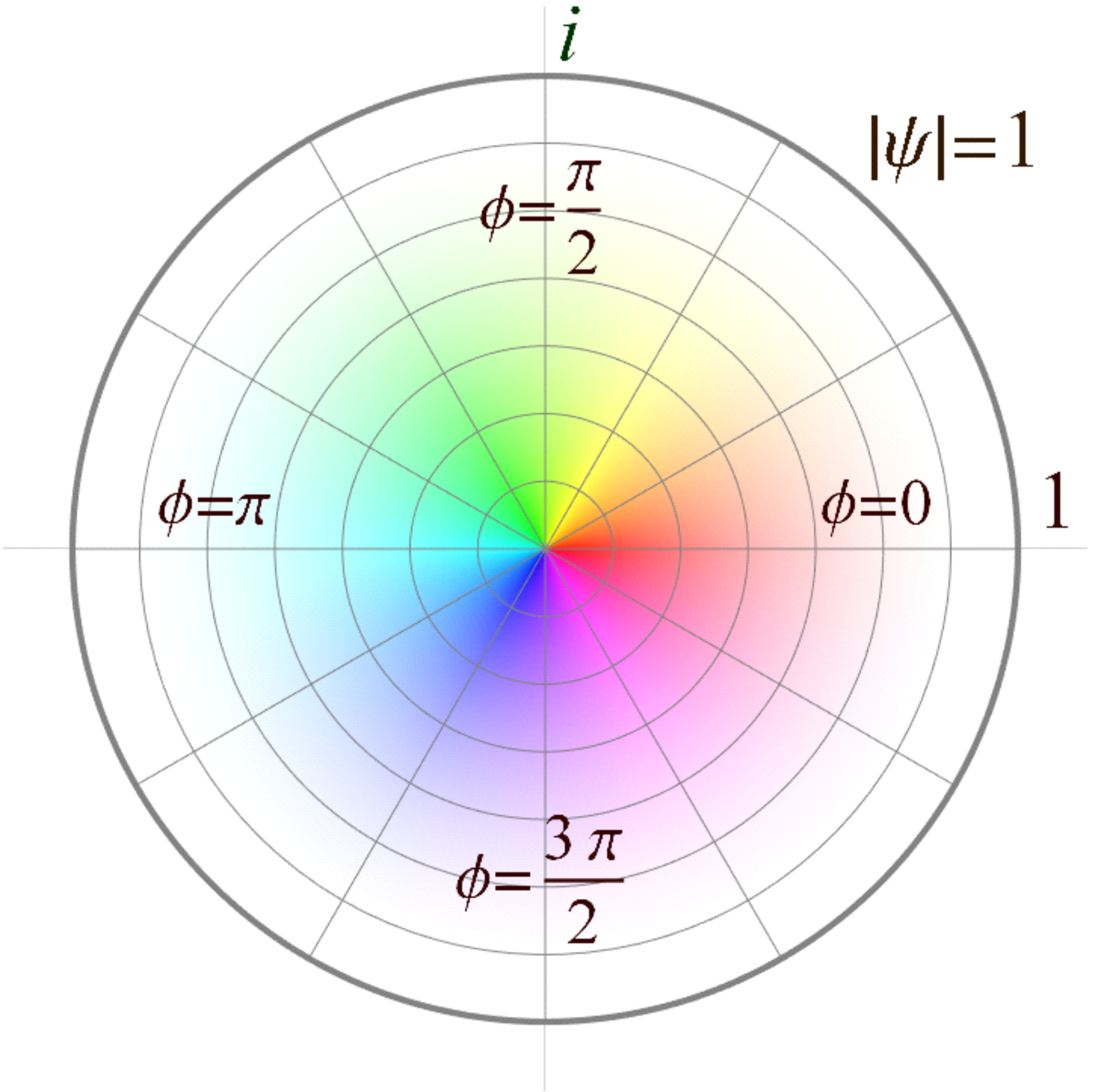}
    \caption{ Color map of the complex plane used for the order parameter visualization. White color stands for a superconducting state $|\psi| \approx 1$ and more saturated colors stand for suppressed superconductivity. The color encodes the order parameter phase. }
    \label{fig:complex_map}
\end{figure}

\begin{figure}
    \centering
        \includegraphics[width=0.96\textwidth]{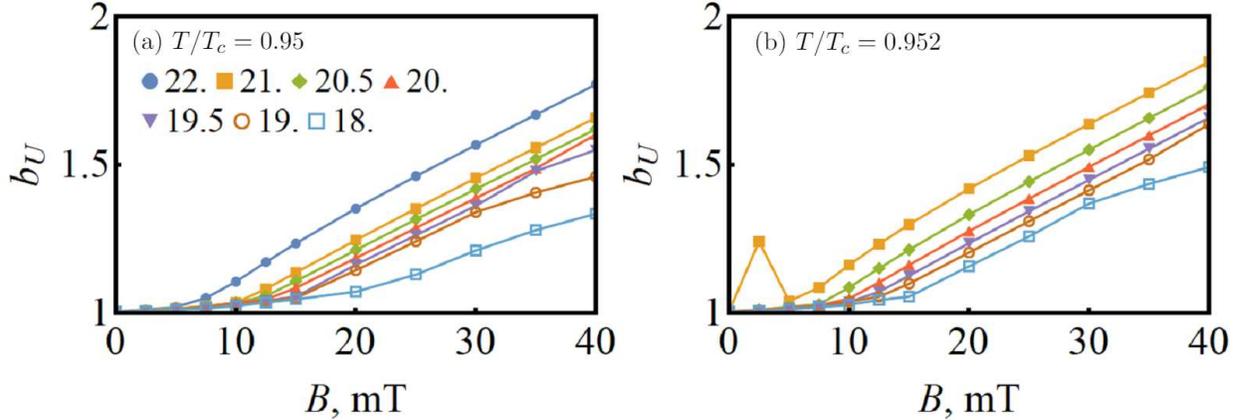}
    \caption{Function $b_U$ of Eq. (\ref{eq:bu}) constructed for plots from Fig. \ref{fig:adiabatic_U}.}
    \label{fig:adiabatic_b_u}
\end{figure}

When the system exerts the regime with vortex chains inside a band of suppressed superconductivity, this band appears in a region with maximal normal magnetic field. We suggest a simple model, considering that a point of the cylinder is in a perfect superconducting state if the normal magnetic field is smaller than some characteristic value $B_1$ (it is a conventional parameter here). Otherwise, it conducts normal current and possesses conductivity $\sigma$. In this case, the fraction of the cylindrical surface, where the normal magnetic field is smaller than $B_1$, is $\text{arccos}(B_{1}/B)$. Consequently, the induced voltage can be estimated as
\begin{equation}
    U_{est} = \frac{4jR}{\sigma} \text{arccos}\frac{B_{1}}{B},
\end{equation}
and the function
\begin{equation}\label{eq:bu}
    b_U = \left(\cos\frac{\sigma U}{4jR}\right)^{-1}
\end{equation}
applied to the numerical results should behave linearly as a function of $B$ in such a regime. Generally, the plots for $T/T_c = 0.95, 0.952$ are indeed piecewisely linear (Fig. \ref{fig:adiabatic_b_u}). Besides this, the plots contain kinks, indicating the regime change. For example, in the plot for $j_{tr}= 19 \text{GA}/\text{m}^2,\,T/T_c=0.95$, there are two kinks at $B=$ 15 and 30 mT. Below 15 mT, the system contains the vortex lattice, in the interval from 15 to 30 mT, the system contains two narrow dense vortex chains with suppressed superconductivity. Above 30 mT, it contains dense vortex chains, decaying into separate vortices.

A transition between the vortex and the phase-slip regimes does not occur when raising the magnetic field (at a constant transport current density), except for the voltage peak at $T/T_c=0.952, \, j_{tr} = 21 \text{ GA}/\text{m}^2$ and $B=2.5 \text{ mT}$.
This peak does not disappear even if the transport current is switched on with $\Delta t_c = 100\text{ ns}$. It disappears at a slightly weaker transport current density $j_{tr} = 20.95 \text{ GA}/\text{m}^2$, and the phase slip appears over the whole analyzed range of the magnetic fields at $j_{tr} = 21.5 \text{ GA}/\text{m}^2$. Therefore, the transition to the phase-slip regime under a gradually switched on transport current occurs only in a narrow window of parameters around the above-mentioned values. In order to describe the mechanism of the appearance of this peak, one can notice that the 2.5 mT magnetic field is strong enough for nucleation of vortices (which have different vorticity in two half-cylinders). On one hand, the magnetic field makes the system state with vortices arranged in the regions with a maximal magnitude of the magnetic field more preferable. On the other hand, the interaction between vortices with different vorticity makes them attract each other. We conjecture if the latter attraction is energetically dominant over the above-described arrangement, then vortices are pushed toward the line opposite to the slit and form a phase slip, which results in a voltage peak.

In the vortex regime, the nucleation and denucleation of moving vortices generate voltage oscillations (typically ranging from 10 to 30 GHz). In low magnetic fields ($\sim$ 2 mT), only several vortices are present at the same instant, what results in a lower frequency and higher amplitude of the voltage oscillations with respect to the average voltage (a larger modulation depth). For example, the system at $T/T_c=0.952, \, j_{tr} = 20 \text{ GA}/\text{m}^2, \, B=2\text{ mT}$ generates a non-harmonic periodic 8.7 GHz signal oscillating with $\sim 10\%$ modulation depth.

\section{Metastable states induced by an abrupt transport current ramping}
The final state of the system may be metastable, depending on the way how the transport current is switched on. For example, the system can arrive either at the phase-slip or at the vortex regime, demonstrating a stable and (quasi-)periodic stationary process, depending on the transport current ramping. In the present section, we analyze the abrupt transport current ramping.

\begin{figure}
    \centering
    \includegraphics[width=0.96\textwidth]{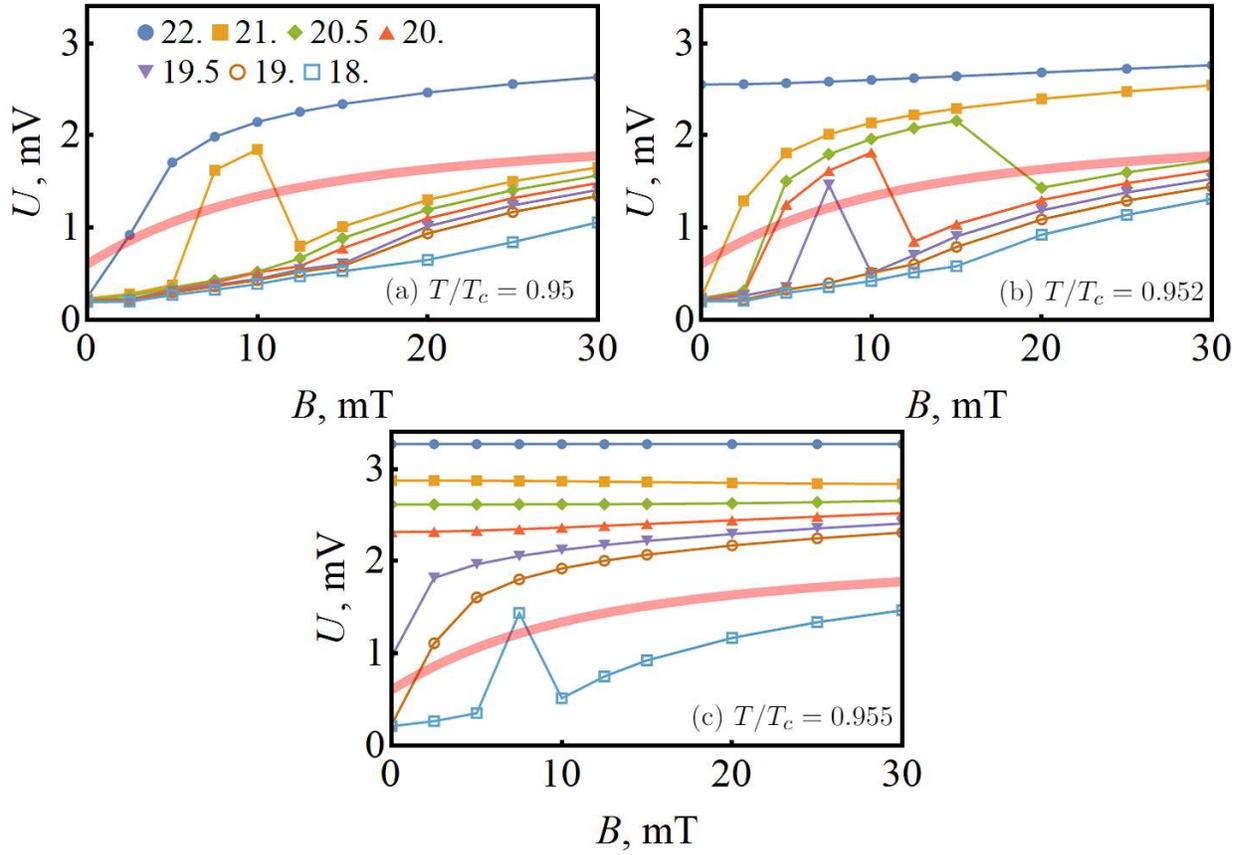}
    \caption{Average voltage induced in Nb open nanotubes with $R=390 \text{ nm}$ and slit $\delta=60 \text{ nm}$ in the external magnetic field $B$ and at a transport current density $j_{tr}$, which is switched on abruptly during $\Delta t_c = 10 \text{ ps}$. Different curves correspond to different values of $j_{tr}$ (in $\text{GA}/\text{m}^2$).  The pink bold line is drawn manually and separates vortex and phase-slip regimes schematically.}
    \label{fig:shock_U}
\end{figure}

\begin{figure}
    \centering
    \includegraphics[width=0.48\textwidth]{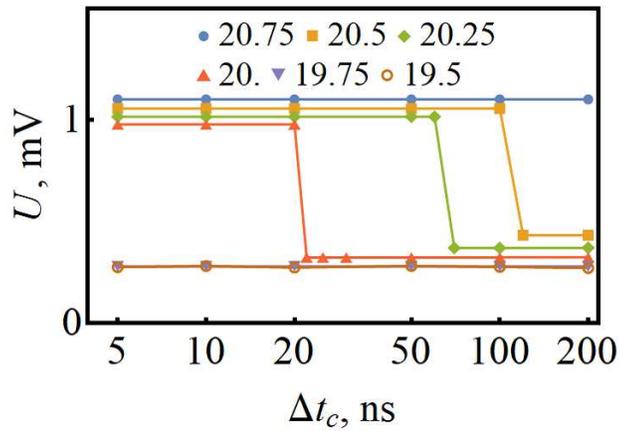}
    \caption{Average voltage induced in Nb open nanotubes with $R=390 \text{ nm}$, $\delta = 60 \text{ nm}$, $T/T_c = 0.95$, $B=10 \text{ mT}$ as a function of the transport current switch-on time $\Delta t_c$ at different values of $j_{tr}$ (in $ 
    \text{GA}/\text{m}^2$).}
    \label{fig:shock_different}
\end{figure}

Reduction of the transport current ramp time to $\Delta t_{c} = 10 \text{ ps}$ leads to the appearance of new peaks in the diagrams (Fig. \ref{fig:shock_U}). The abruptness of the ramping, which is needed to get a phase slip, is smaller if one approaches the critical current density (see example for $T/T_c=0.95$, $B=10 \text{ mT}$ in Fig. \ref{fig:shock_different}). These peaks correspond to the phase-slip regime induced by an abrupt transport current ramping, while in the case of a gradual current switch-on there are two vortex chains. To demonstrate that the phase slip triggered by the abrupt transport current ramping is stable in time, we obtained it for the $T/T_c = 0.952, B=10 \text{ mT}, j_{tr}=20\text{ GA}/\text{m}^2, \Delta t_{c} = 20\text{ ps}$ during 60 ns.

\begin{figure}
    \centering
        \includegraphics[width=\textwidth]{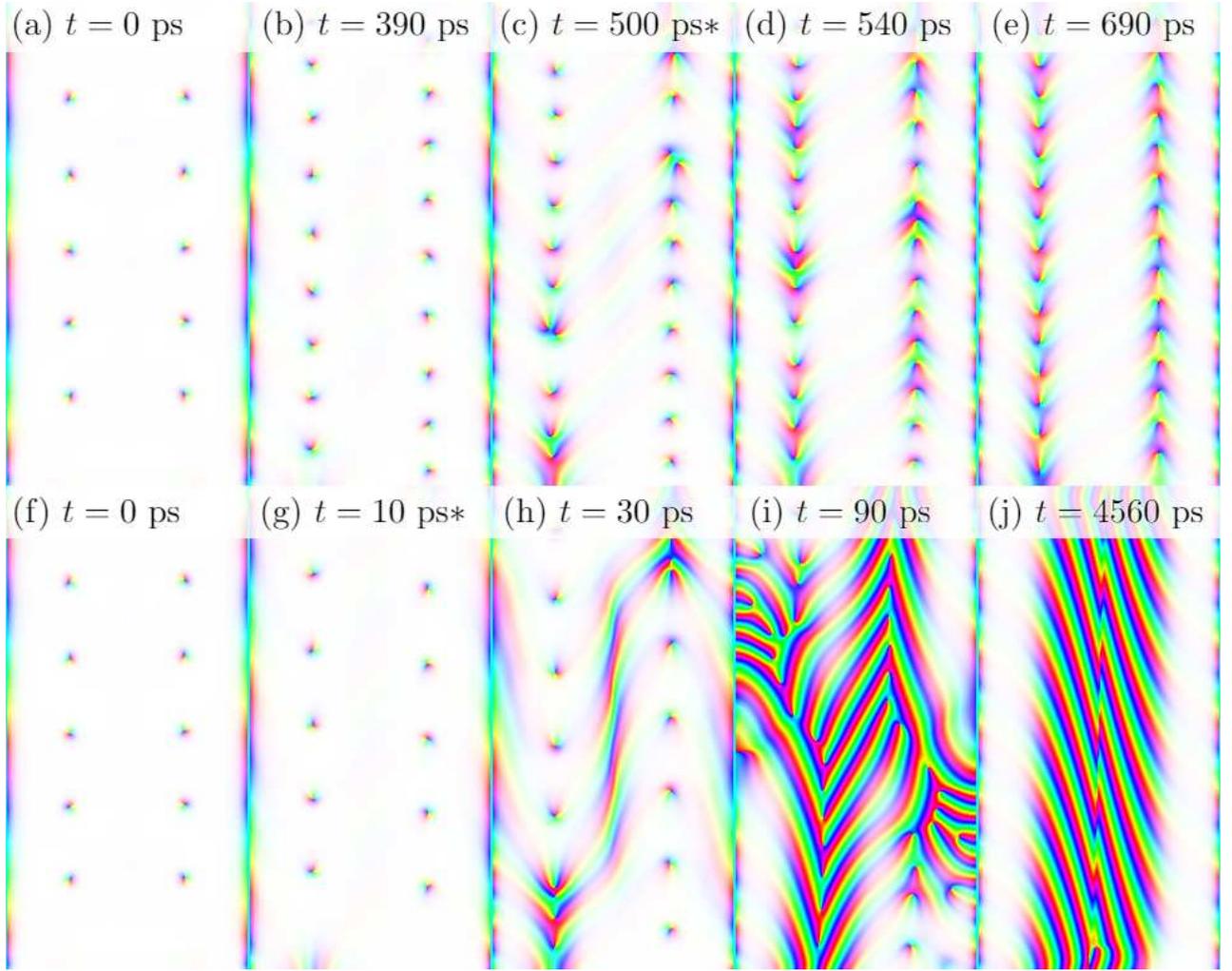}
    \caption{Order parameter evolution for $R=390 \text{nm}$, $T/T_c = 0.952$, $B=10\text{ mT}$, $j_{tr}=20.5 \text{ GA}/\text{m}^2$ at different transport current ramp times $\Delta t_c$: gradual $\Delta t_c=500\text{ ps}$ (upper) and abrupt $\Delta t_c=10\text{ ps}$ (lower). Each panel represents the unfolded open nanotube with height $L=5\,\mu\text{m}$ and the membrane width $W = 2.39 \,\mu\text{m}$. Time $t$ is counted from the moment when the transport current begins to switch on. The asterisk denotes the frame, when the transport current density achieves its final value (at Figs. \ref{fig:different_evolution}c and \ref{fig:different_evolution}g).}
    \label{fig:different_evolution}
\end{figure}

The difference between the evolution at gradual and abrupt transport current ramping is depicted in Fig. \ref{fig:different_evolution}. In the gradual current switch-on case, the system has time to adapt to a higher transport current density, so that new vortices nucleate in both half-cylinders. Finally, it ends up with two dense vortex patterns. In the abrupt current switch-on case, the system does not manage to nucleate new vortices fast enough in response to the transport current growth. This creates conditions for the suppression of the superconductivity in the region opposite to the cylinder slit. The region with suppressed superconductivity grows and captures the vortices from the half-cylinders. Vortex chains from both half-cylinders begin moving toward the line opposite to the cylinder slit, and finally, they come up with a vortex-antivortex chain constituting the phase slip. In the intermediate case, if the abruptness is not enough to create a phase slip, the region of the suppressed superconductivity does not grow, but rather decays into several vortex-antivortex pairs, which join the vortex chains in the corresponding half-cylinders.

\section{Metastable states induced by abrupt magnetic field ramping}

\begin{figure}
    \centering
    \includegraphics[width=0.96\textwidth]{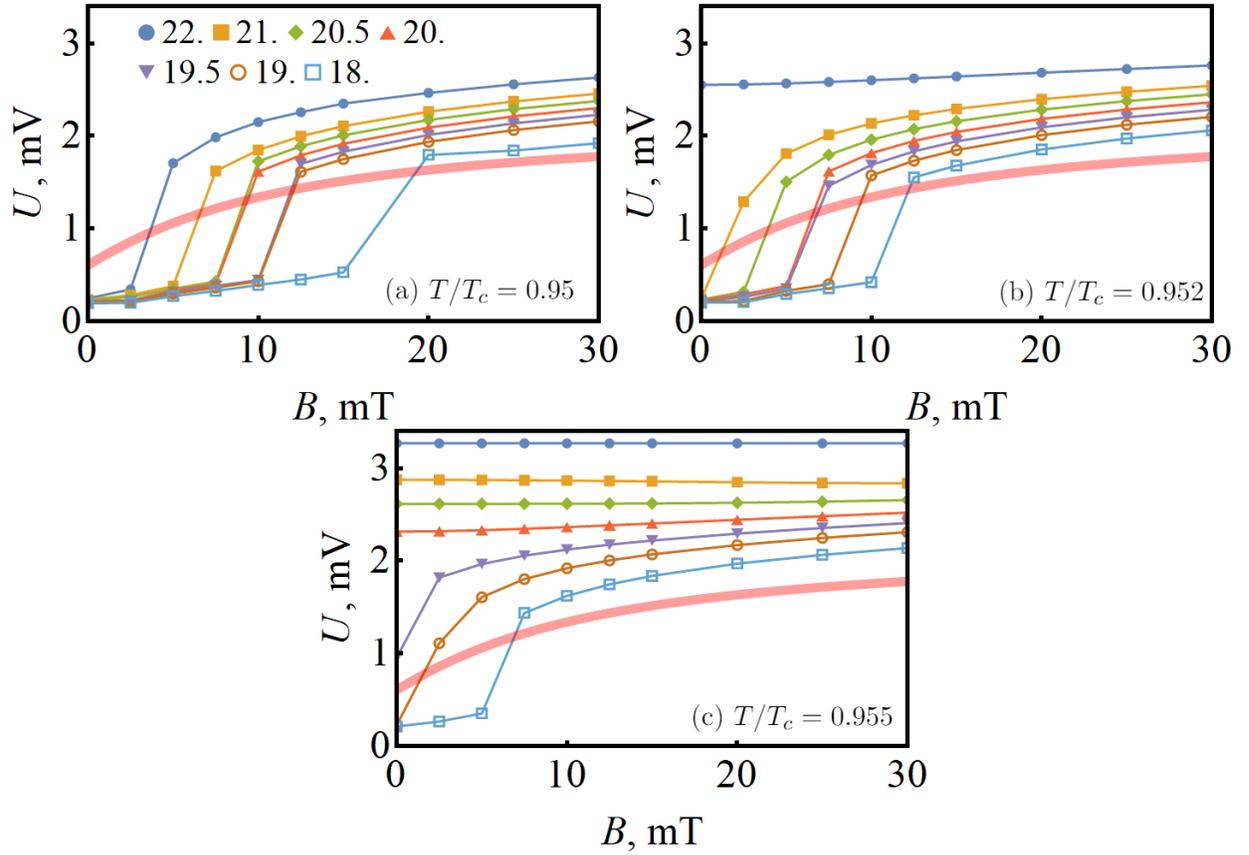}
    \caption{Average voltage induced in Nb open nanotubes with $R=390 \text{ nm}$ and slit $\delta = 60 \text{ nm}$, where first the transport current is switched on and then abruptly the magnetic field $B$. The magnetic field ramp time is $\Delta t_f = 10 \text{ ps}$. Different curves correspond to different values of the transport current density $j_{tr}$ (in $ 
    \text{GA}/\text{m}^2$). The pink bold line is drawn manually and separates vortex and phase-slip regimes schematically.}
    \label{fig:abrupt_magnetic_U}
\end{figure}

In this section, we demonstrate that a similar effect can be achieved if first the transport current and then the magnetic field are switched on. The calculations with gradual ramping (both magnetic field and current switch-on slope) exhibited the results similar to those discussed in Sec. \ref{sec:gradual}. The abrupt magnetic field ramping (magnetic field ramp time is $\Delta t_f = 10\text{ ps}$) gives rise to a transition from the vortex to the phase-slip regime (Fig. \ref{fig:abrupt_magnetic_U}).

\begin{figure}
    \centering
    \includegraphics[width=0.95\textwidth]{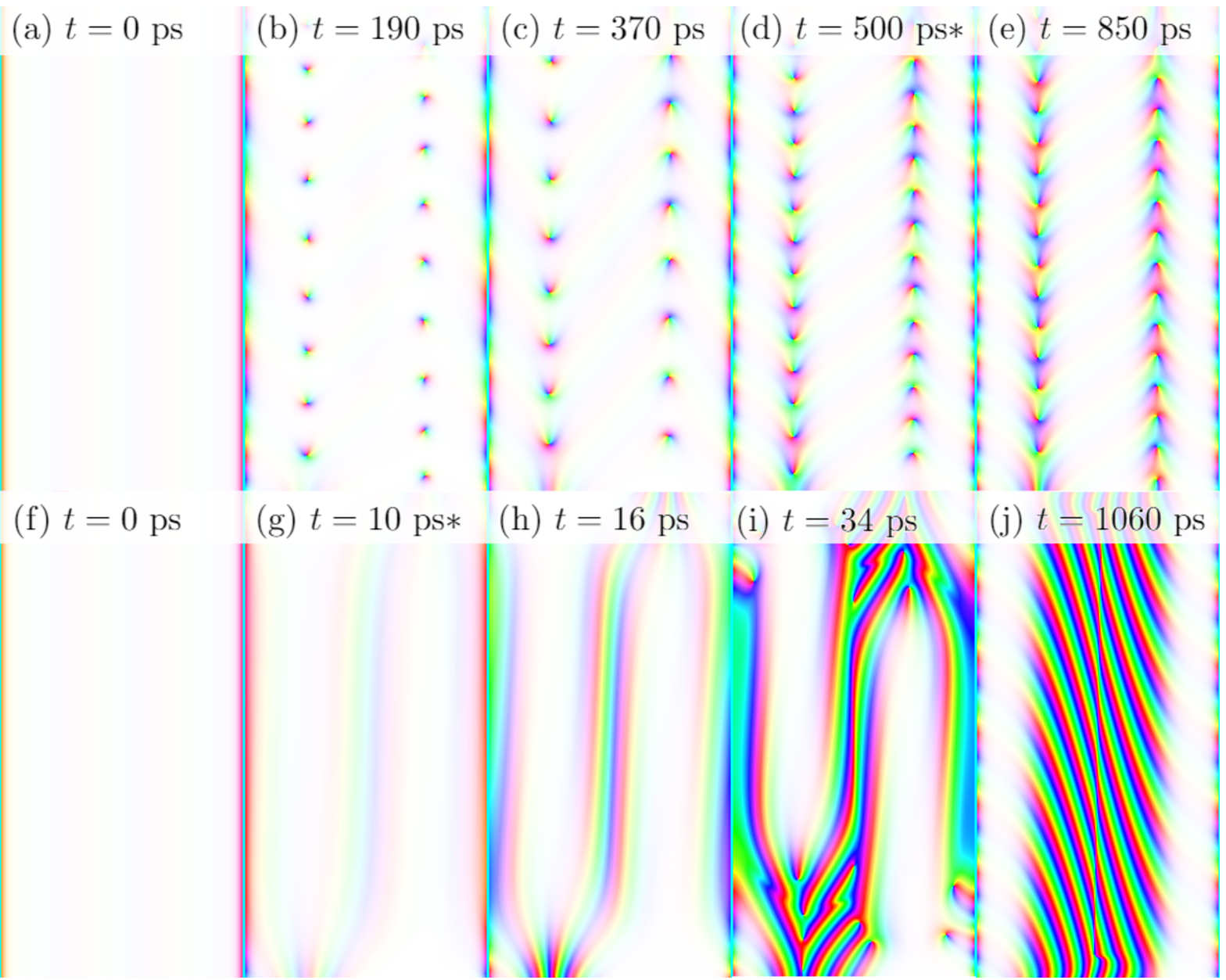}
    \caption{Order parameter evolution for $R=390 \text{ nm}$, $T/T_c = 0.952$, $B=10 \text{ mT}$, $j_{tr}=20.5 \text{ GA}/\text{m}^2$ for different magnetic field ramp times $\Delta t_f$: gradual $\Delta t_f = 500 \text{ ps}$ (upper) and abrupt $\Delta t_f = 10 \text{ ps}$ (lower). Each panel represents the unfolded open nanotube with height $L=5\,\mu\text{m}$ and the membrane width $W = 2.39 \,\mu\text{m}$. Time $t$ is counted from the moment when the magnetic field begins to switch on. The asterisk denotes the frame, when the magnetic field achieves its final value (at Figs. \ref{fig:different_evolution_magnetic}d and \ref{fig:different_evolution_magnetic}g).}
    \label{fig:different_evolution_magnetic}
\end{figure}

If the current is switched on first, a gradual magnetic field ramping leads to the vortex regime and an abrupt magnetic field ramping leads to the phase slip depicted in Fig. \ref{fig:different_evolution_magnetic}. As the current density is subcritical, the state before the magnetic field is switched on does not contain any topological defects. In the case of a gradual magnetic field switch-on, a vortex chain nucleates in response to the magnetic field change. In the case of an abrupt magnetic field switch-on, the system does not manage to generate vortices as fastly as required by the raising magnetic flux, and the superconductivity is suppressed in the region opposite to the slit (in the middle of the panel Fig. \ref{fig:different_evolution_magnetic}h). In the region of suppressed superconductivity, vortices and antivortices nucleate, leading to the phase-slip regime. 

\section{Hysteresis in the current-voltage characteristic controlled by the magnetic field}
As demonstrated in the previous sections, depending on the way how the current or magnetic field are switched on, the system can arrive at different (meta)stable states for the same set of the final parameters. This means that there is an appreciable energy barrier between different regimes of superconducting dynamics. As a result, one can expect a hysteresis effect. To demonstrate the hysteresis effect, the following steps are performed:
\begin{itemize}
    \item H1. Prepare an initial random state (in the same way as in S1).
    \item H2. Switch on the magnetic field linearly from 0 to its final value $B$ during 10 ps and let the system relax during 100 ps.
    \item H3. Switch on the transport current density linearly from 0 to 14 GA/m${}^2$ during 100 ps and let the system relax during 400 ps.
    \item H4. Increase $j_{tr}$ by 0.129 GA/m${}^2$ during 100 ps and let the system relax during 400 ps. Repeat it until $j_{tr}$ reaches the value of 23 GA/m${}^2$. 
    \item H5. Repeat H4 in the opposite direction: decrease the transport current density from 23 to 14 GA/m${}^2$.
\end{itemize}

\begin{figure}
    \centering
    \includegraphics[width=0.55\textwidth]{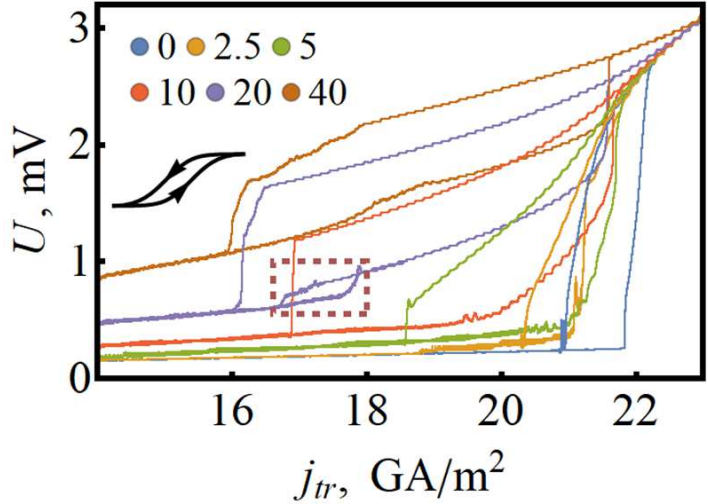}
    \caption{The hysteresis effect in the current-voltage characteristic of Nb open nanotubes with $R=390\text{ nm}$, $\delta = 60 \text{ nm}$, $T/T_c = 0.952$ for different values of the magnetic field (indicated in mT near the corresponding marks). The scheme in the inset shows the sequence of the voltage changes during the steps H4 and H5.}
    \label{fig:hysteresis}
\end{figure}

The calculations are conducted for $T/T_c=0.952$ at different values of $B$ (Fig. \ref{fig:hysteresis}). The hysteresis effect is practically negligible at $B=0, 2.5 \text{ mT}$. Actually, the visible hysteresis effect for these values appears due to the short relaxation time (which is a computational restriction). Hence, there is no energy barrier between the phase-slip and the vortex regimes for the weak magnetic field. Starting from $\sim$5 mT, the hysteresis effect is distinctly manifested. The hysteresis loop is wider along $j_{tr}$ and narrower along $U$ for higher magnetic fields. The hysteresis loop is wider along the $j_{tr}$ direction and smaller along the $U$ direction for stronger magnetic fields. As a result the hysteresis loop area has maximum at some value of the magnetic field $B$ between 10 mT and 20 mT. The hysteresis loop area can be understood as a power of excessive dissipation attributed to the existence of the higher branch of the loop in comparison with the process if only the lower branch would exist. In addition to the hysteresis effect, which is the result of transitions between the vortex and the phase-slip regimes, we have found a small hysteresis loop (in the burgundy frame, Fig. \ref{fig:hysteresis}) for the transition between a sparse vortex lattice (Fig. \ref{fig:adiabatic_regimes}a) and a dense vortex chain (Fig. \ref{fig:adiabatic_regimes}b).

\section{Discussions}

We have shown that for the gradual switch-on of the magnetic field and/or transport current in the superconductor open nanotubes, the transition between the vortex and the phase-slip regimes depends on the magnetic field only weakly. Depending on the switch-on, the superconductor can arrive at different (meta)stable states for the same final external parameters (transport current density and magnetic field). Abrupt magnetic-field or transport-current ramping can lead to the transition to the phase-slip regime. The ramp time of the external parameters can be used to control the vortex/phase-slip transition in the nanotubes. The energy barrier between phase-slip and vortex states occurs to be higher in stronger magnetic fields. This barrier leads to the existence of the appreciable hysteresis effect in the current-voltage characteristics controlled by the magnetic field. The system state near the critical transport current density depends on the way by which the final external parameters (transport current density and magnetic field) are achieved, constituting the {\it memory effect}. Dependency on the switch-on regime and the hysteresis effect appears dynamically (due to the energy barrier between different superconducting regimes) with no contribution of the heat-related effects. Additionally, we pay attention to the importance of taking into account the realistic switch-on regimes of the external parameters (e.g., transport current and external magnetic field) in numerical calculations.

The numerical simulation of superconducting dynamics demonstrates the voltage peak even for the gradual evolution at $T/T_c=0.952, \, B=2.5\text{ mT}, \, j_{tr}=21\text{ GA}/\text{m}^2$. This peak can be observed in a narrow window of parameters around the above-mentioned values. We conjecture that it appears when the magnetic field is strong enough to begin nucleating vortices, but the current-induced attraction between vortices in two half-cylinders is still energetically more preferable, than the arrangement of vortices in the regions with a maximal magnitude of the magnetic field. Moreover, in the considered system within the region of parameters, where there are only a few vortices, 8.7 GHz alternating voltage is generated with a relatively large modulation depth $\sim10\%$.

The presented new results shed light on the interplay between different regimes of the superconducting dynamics in open nanotubes with clear perspectives for application in various fields of nanotechnology. Understanding of the order parameter transition between the corresponding regimes helps manufacture nanosensors of the magnetic field and quantum-interference-based filters and switchers. The existence of the metastable states is promising to boost the progress in fluxon-based devices for quantum computing and to improve the performance of the nanostructured bolometers and THz-detectors.

\begin{acknowledgments}
This work has been supported by the German Research Foundation (DFG) project \#FO 956/6-1 (Germany) and European Cooperation in Science and Technology—COST Action \#CA16218 (NANOCOHYBRI) under the Virtual Networking Grant E-COST-GRANT-CA16218-1f6dba72 ``Topology- and geometry-driven transport properties of self-rolled superconductor nanoarchitectures''. The authors are grateful to Dmitri V. Gal’tsov for his advice and support of the project and to Victor Ciobu for the technical support. V.M.F. acknowledges partial support from the MEPhI (Russia) and the ZIH TU Dresden (Germany) for providing its facilities for high throughput calculations.
\end{acknowledgments}

\end{document}